%% file: 0_main.tex
\documentclass[conference]{IEEEtran}
\usepackage[caption=false,font=normalsize,labelfont=sf,textfont=sf]{subfig}
\usepackage{cite}
\usepackage{amsmath}		
\usepackage{amsfonts}		
\usepackage{amssymb}        
\usepackage{ulem}
\ifCLASSINFOpdf
\usepackage{graphicx}
\usepackage{txfonts}
\else
\usepackage{graphicx}
\fi
\usepackage{float}
\usepackage{multicol}
\usepackage{epstopdf}
\usepackage{breqn}
\usepackage{multirow}
\usepackage{caption}
\usepackage{soul}
\usepackage{algorithm}
\usepackage{algpseudocode}
\usepackage{mathtools}
\usepackage{xcolor}

\usepackage{bbm}
\usepackage{dblfloatfix}

\usepackage{hyperref}
\hypersetup{
    colorlinks=true,
    linkcolor=blue,
    filecolor=magenta,      
    urlcolor=blue,
}
\usepackage{acro}
\input{acronyms}

\begin{document}

\title{Low Complexity High Speed Deep Neural Network Augmented Wireless Channel Estimation}

\author{\IEEEauthorblockN{Syed Asrar ul haq, Varun Singh, Bhanu Teja Tanaji and Sumit Darak}
 \IEEEauthorblockA{Electronics and Communications Department, IIIT-Delhi, India-110020 \\e-mail: \{syedh, varun22189, bhanu22155, sumit\}@iiitd.ac.in}
}

\maketitle

\input{1_abstract}
\input{2_introduction}

\input{3_system_model}
\input{4_channel_estimation}
\input{5_soc_implementation}

\input{6_results}

\input{7_conclusion}

\bibliography{References.bib}
\bibliographystyle{ieeetr}
\end{document}

%% file: acronyms.tex
\DeclareAcronym{fpga}{
  short=FPGA,
  long=Field Programmable Gate Array,
}

\DeclareAcronym{soc}{
  short=SoC,
  long=System on Chip,
}

\DeclareAcronym{phy}{
  short=PHY,
  long=physical layer,
}

\DeclareAcronym{dl}{
  short=DL,
  long=deep learning,
}

\DeclareAcronym{dnn}{
  short=DNN,
  long=deep neural network,
}
\DeclareAcronym{lsdnn}{
  short=LSDNN,
  long=least square deep neural network,
}

\DeclareAcronym{cnn}{
  short=CNN,
  long=Convolutional Neural Network,
}
\DeclareAcronym{csi}{
  short=CSI,
  long=channel state information,
}

\DeclareAcronym{ls}{
  short=LS,
  long=least squares,
}

\DeclareAcronym{mmse}{
  short=MMSE,
  long=minimum mean square error,
}

\DeclareAcronym{mse}{
  short=MSE,
  long=mean square error,
}

\DeclareAcronym{ofdm}{
  short=OFDM,
  long=Orthogonal Frequency Division Multiplexing,
}

\DeclareAcronym{srcnn}{
  short=SRCNN,
  long=Super-Resolution CNN,
}
\DeclareAcronym{dncnn}{
  short=DnCNN,
  long=Denoising CNN,
}

\DeclareAcronym{snr}{
  short=SNR,
  long=signal-to-noise ratio,
}

\DeclareAcronym{iresnet}{
  short=iResNet,
  long=Interpolation ResNet,
}
\DeclareAcronym{resnet}{
  short=ResNet,
  long=Residual Neural Network,
}

\DeclareAcronym{fcnns}{
  short=FCNNs,
  long=Fully Connected Neural Networks,
}

\DeclareAcronym{fcnn}{
  short=FCNN,
  long=Fully Connected Neural Network,
}

\DeclareAcronym{sta}{
  short=STA,
  long=Spectral Temporal Averaging,
}
\DeclareAcronym{ifft}{
  short=IFFT,
  long=Inverse Fast Fourier Transform,
}
\DeclareAcronym{lts}{
  short=LTS,
  long=Long Training Sequence,
}

\DeclareAcronym{trfi}{
  short=TRFI,
  long=Time domain Reliability test Frequency domain Interpolation,
}
 
\DeclareAcronym{rnns}{
  short=RNNS,
  long=Recurrent Neural Networks,
}

\DeclareAcronym{rnn}{
  short=RNN,
  long=Recurrent Neural Network,
}

\DeclareAcronym{lstm}{
  short=LSTM,
  long=Long Short-Term Memory,
}

\DeclareAcronym{gru}{
  short=GRU,
  long=Gated Recurrent Unit,
}

\DeclareAcronym{awgn}{
  short=AWGN,
  long=Additive White Gaussian Noise,
}

\DeclareAcronym{ps}{
  short=PS,
  long=processing system,
}
 
\DeclareAcronym{pl}{
  short=PL,
  long=programmable logic,
}

\DeclareAcronym{pe}{
  short=PE,
  long=processing element,
}

\DeclareAcronym{mac}{
  short=MAC,
  long=Multiply-Accumulate,
}

\DeclareAcronym{tdl}{
  short=TDL,
  long=tapped delay line,
}

\DeclareAcronym{vtv}{
  short=VTV,
  long=vehicle-to-vehicle,
}

\DeclareAcronym{rtv}{
  short=RTV,
  long=roadside-to-vehicle,
}

\DeclareAcronym{nmse}{
  short=NMSE,
  long=normalized mean squared error,
}

\DeclareAcronym{ber}{
  short=BER,
  long=bit error rate,
}

\DeclareAcronym{relu}{
  short=ReLU,
  long=Rectified Linear Unit,
}

\DeclareAcronym{bpsk}{
  short=BPSK,
  long=Binary Phase Shift Keying,
}

\DeclareAcronym{sbs}{
  short=SBS,
  long=Symbol by Symbol,
}

%% file: 1_abstract.tex
\begin{abstract}
The channel estimation (CE) in wireless receivers is one of the most critical and computationally complex signal processing operations. 
Recently, various works have shown that the deep learning (DL) based CE outperforms conventional minimum mean square error (MMSE) based CE, and it is hardware-friendly. 
However, DL-based CE has higher complexity and latency than popularly used least square (LS) based CE. In this work, we propose a novel low complexity high-speed Deep Neural Network-Augmented Least Square (LC-LSDNN) algorithm for IEEE 802.11p wireless physical layer and efficiently implement it on Zynq system on chip (ZSoC). 
The novelty of the LC-LSDNN is to use different DNNs for real and imaginary values of received complex symbols. This helps reduce the size of DL by 59\% and optimize the critical path, allowing it to operate at 60\% higher clock frequency. We also explore three different architectures for MMSE-based CE. We show that LC-LSDNN significantly outperforms MMSE and state-of-the-art DL-based CE for a wide range of signal-to-noise ratios (SNR) and different wireless channels. Also, it is computationally efficient, with around 50\% lower resources than existing DL-based CE.

\end{abstract}

%% file: 2_introduction.tex
\section{Introduction} \label{sec:introduction}


The wireless physical layer (PHY) enables efficient transmission of digital information from transmitter to receiver in the presence of fading channels and hardware impairments. With the evolution of \ac{dl}, various works have explored its usefulness for wireless PHY \cite{end_to_end1,attention1,animesh,Himani_DL,DL_Rohith,deepwiphy,channelSounding2,lstmdpata,lsdnn}. In \cite{end_to_end1}, wireless PHY is replaced with single DL model and has shown to offer better performance and adaptability in unknown channel environment. However, a single DL-based PHY is incompatible with existing standards due to the need for intermediate control signal communications between transmitters and receivers to allow multiple users to coexist in wireless networks. The alternative standard-compatible approach is to replace or augment one or more signal processing operations with DL model \cite{deepwiphy,DL_Rohith,channelSounding2}.

Channel estimation (CE) is a critical and computationally intensive task. It estimates \ac{csi} at the receiver using known transmitted data. Wireless receivers utilize \ac{csi} for reliable data reception, while transmitters employ it to select appropriate channel and PHY parameters. Additionally, \ac{csi} enhances PHY security in wireless networks and improves the accuracy of indoor localization techniques. The widely used least square (LS) based CE is simple and easy to implement but performs poorly at low signal-to-noise ratio (SNR). Another statistical method is minimum mean square error (MMSE) based CE, which offers improved performance but needs computationally intensive matrix inverse operation.

The CE approach depends on wireless PHY. In cellular PHY, CE uses pilots embedded in the data frame, while in IEEE 802.11 PHY, CE is done using a preamble. In this work, we focus on preamble-based CE for IEEE 802.11p PHY. Without loss of generality, the proposed approach can be extended to cellular PHY with additional interpolation operations~\cite{attention1,animesh}. For preamble-based PHY,  \ac{dl} is employed to enhance the estimation of each \ac{ofdm} symbol either as pre-processing \cite{lstmdpata}, or as postprocessing~\cite{lsdnn}. 
While \ac{dl}-based CE has demonstrated improved functional accuracy, limited research has been conducted on mapping of \ac{dl}-based CE on \ac{soc} \cite{lsdnn}. This is critical since wireless PHY must be deployed on edge platforms, and hence, constraints of such platforms must be taken into account in algorithm design \cite{dlinphy}.

In this work, we aim to design low-complexity, high-speed DL-based CE and efficiently implement it on SoC. The contributions of the paper are summarized as follows:

\begin{enumerate}
    \item We propose a novel low complexity high-speed Deep Neural Network-Augmented Least Square (LC-LSDNN) algorithm for the IEEE 802.11p wireless PHY. The novelty of the LC-LSDNN is to use distinct DNNs for real and imaginary components of received complex symbols, reducing DNN size and optimizing the critical path for operation at higher clock frequencies.

\item We assess CE performance in end-to-end PHY across various SNRs and wireless channels. We show that the proposed LC-LSDNN outperforms existing approaches. Furthermore, we demonstrate that the DL models can be trained using a high SNR dataset instead of an ideal channel model, making it practically useful.

\item We efficiently map the LC-LSDNN on Zynq ZC706 SoC. We explore three different architectures for MMSE-based CE and optimize each architecture for the lowest word length (WL) and maximum feasible clock frequency. We show that the LC-LSDNN requires fewer resources, lower power, and lower latency than state-of-the-art CE approaches.

\end{enumerate}

The paper is organized as follows: Section \ref{sec:sysModel} discusses the system model and reviews existing CE approaches. In Section \ref{sec:proposed}, we discussed the proposed LC-LSDNN algorithm, simulation, and complexity results, followed by architecture details in Section \ref{sec:socImplementation} and implementation results in Section \ref{sec:results}. Section \ref{sec:conclusions} concludes the paper. 



%% file: 3_system_model.tex
\section{System Model and Review of Wireless Channel Estimation Approaches} \label{sec:sysModel}

\begin{figure}[!t]
\centering
\includegraphics[scale=0.55]{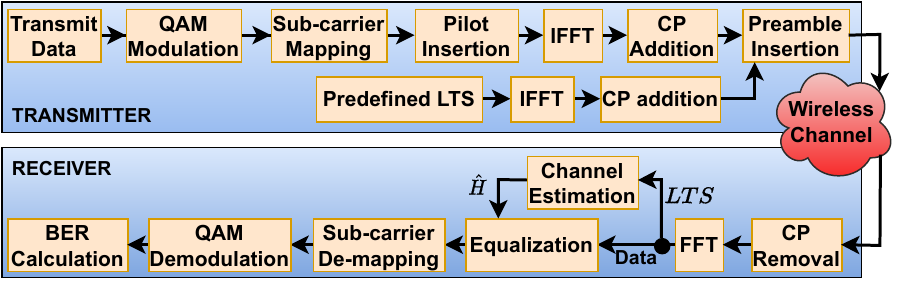}
\caption{\small  Building blocks of IEEE 802.11p OFDM transceiver PHY.}
    \label{fig:BlockDiagram}
\end{figure}
\begin{figure}[!t]
\vspace{-0.2cm}
\centering
\includegraphics[scale=0.7]{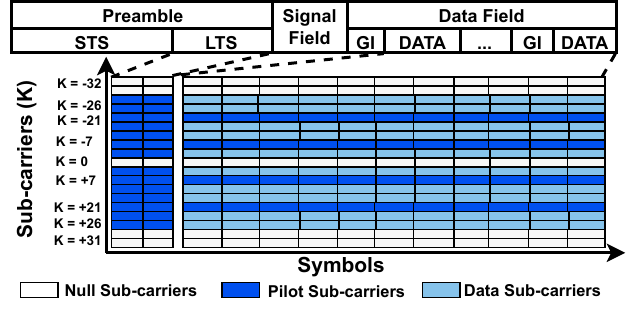}
\caption{\small  Frame structure of IEEE 802.11p  OFDM transceiver PHY.}
    \label{fig:frameStruct}
    \vspace{-0.2cm}
\end{figure}

We consider an OFDM-based IEEE 802.11p standard wireless PHY as shown in Fig.~\ref{fig:BlockDiagram}. It consists of a data frame comprising a preamble followed by the data field as shown in Fig.~\ref{fig:frameStruct}. The preamble consists of ten short training symbols (STS) for signal detection and timing synchronization and two long training symbols (LTS) for CE. The preamble is OFDM modulated, employing 64 sub-carriers. Among these, 52 are allocated to LTS, while the remaining are null sub-carriers. The information bits to be transmitted are subjected to data modulation using quadrature amplitude modulation  (QAM) followed by OFDM modulation. Of 64 sub-carriers, 48 are used for data, 12 are null subcarriers, and 4 are pilots for carrier offset estimation at the receiver.

At the receiver, OFDM demodulation is performed, followed by CE using LTS. The estimated channel is used to equalize the received data subcarriers, followed by data demodulation. 
We use bit error rate (BER) as an end-to-end performance metric and normalized mean square error (NMSE) as a performance metric for estimation performance.



%% file: 4_channel_estimation.tex
When a transmitted signal traverses a wireless channel, it undergoes fading due to multipath propagation and experiences a Doppler shift resulting from mobility between transmitter and receiver. This work assumes a low mobility scenario where the channel estimated using the preamble remains unchanged for the entire data frame. However, the proposed work can be extended to high mobility scenarios using data-pilot aided (DPA) approaches where channel estimated using a preamble can be updated further using real and virtual pilots \cite{lstmdpata}. Next, we review various CE approaches.


\vspace{-0.25cm}
\subsection{Statistical Channel Estimation Approaches}
If $X \in \mathbb{C}^{K}$ represents the transmitted OFDM symbol with $K$ subcarriers, $H \in \mathbb{C}^{K}$ denotes the channel gain vector for the current \ac{ofdm} symbol, and $Z \in \mathbb{C}^{K}$ represents the \ac{awgn}, then the received signal $Y \in \mathbb{C}^{K}$ can be expressed as follows:
\begin{equation}
    Y = H \odot X + Z
\end{equation}
where $\odot$ is the element-wise multiplication operator. 
The LS estimate aims to minimize the squared error between the received signal and actual channel response.
In an \ac{ofdm} PHY with circular convolution due to CP, $\hat{H}_{LS}[i,k]$ for subcarrier $k$ of symbol $i$ is simplified as follows:
 \begin{equation}
     \hat{H}_{LS}[i,k] = \frac{Y[i,k]}{X[i,k]}
 \end{equation}
The channel estimates are calculated for both LTS, and averaged to get the final estimate.
 LS estimation is simple and easy to implement, but it performs poorly at low SNRs due to its lack of consideration for the influence of noise. MMSE leverages prior knowledge of channel and noise statistics to improve the LS. It is given as
 \begin{equation} \label{eqn:mmse}
     \hat{H}_{MMSE} = R_{HH} \left ( R_{HH} + \frac{K \mathcal{N}_0}{E_p} I \right)^{-1} \hat{H}_{LS}
 \end{equation}
As discussed in Section~\ref{sec:results}, MMSE offers improved performance but computationally complex. 

\subsection{DL Augmented Channel Estimation}
DL is currently being explored for CE due to its exceptional feature extraction capabilities from raw data, enabling it to comprehend and generate meaningful representations \cite{dlForPhy}. DL architectures for inference are hardware-friendly due to simple arithmetic operations.
As discussed in Section~\ref{sec:introduction}, we focus on low-complexity DL-augmented CE compared to existing approaches where CE is replaced with large-size DL, resulting in a high area, power, and delay penalty. In \cite{lsdnn}, a \ac{dnn}-augmented CE is discussed, followed by its realization on ZSoC. 
Though it outperforms LS and MMSE, it has high computation complexity and latency compared to LS. We aim to address these challenges by novel contributions at the algorithm and architecture levels.


\section{Proposed DL Augmented Channel Estimation} \label{sec:proposed}
In this section, we present the proposed low complexity least square augmented DNN-based CE (LC-LSDNN) followed by performance analysis using floating-point arithmetic and theoretical complexity analysis with existing DL-based CE. 

\begin{figure}[!b]
 \vspace{-0.2cm}
\centering
\includegraphics[scale=0.7]{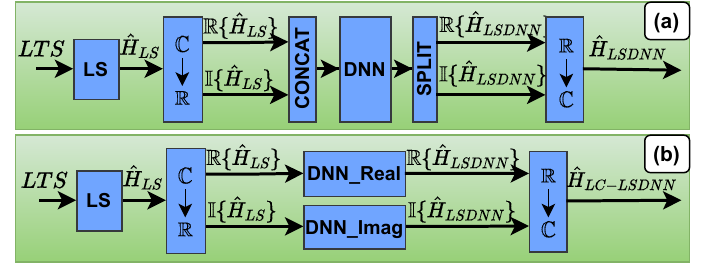}
\caption{\small  DL-based CE in (a) \cite{lsdnn}, and (b) Proposed LC-LSDNN.}
    \label{fig:LSDNN}
    \vspace{-0.2cm}
\end{figure}

\subsection{Proposed LC-LSDNN}

In 802.11p PHY, LS-based CE produces a complex output, $\hat{H}_{LS} \in \mathbb{C}^{k}$. Existing methods \cite{lsdnn,deepwiphy} concatenate real and imaginary components for DNN processing, as shown in Fig.~\ref{fig:LSDNN}(a). However, this approach has high memory and compute complexity. Recent work \cite{channelSounding2} explores a more efficient strategy using separate neural networks for real and imaginary components. Though this approach seems promising, an in-depth analysis of its functionality and complexity on floating and fixed-point architectures has not been done yet.

The proposed approach employs two distinct DNNs to process the real and imaginary components independently, as illustrated in Fig.~\ref{fig:LSDNN}(b). 
The LS estimate, denoted as $\hat{H}_{LS} \in \mathbb{C}^{k_{on}}$, is initially decomposed into its real $\Re\{\hat{H}_{LS} \} \in \mathbb{R}^{k_{on}}$ and imaginary component $\Im\{\hat{H}_{LS} \} \in \mathbb{R}^{k_{on}}$, where $k_{on}$ are the active subcarriers. These components are individually fed into their respective trained DNNs, producing outputs that correspond to the real and imaginary components of the channel estimates.
The outputs of these two DNNs are combined, resulting in complex-valued channel estimates $\hat{H}_{LC-LSDNN} \in \mathbb{C}^{k_{on}}$.

The LC-LSDNN consists of a \ac{fcnn} with a single hidden layer utilizing ReLU activation function and an output layer with a linear activation function.
The input and output layer is of size $k_{on}$ while
the hidden layer size is $k_{on}/2$.

\begin{table}[!b]
\centering
\caption{\small DNN parameters.}
\label{tab:dnnParam}
 \resizebox{\columnwidth}{!}{
    \begin{tabular}{|l|l|}
        \hline
        \textbf{Parameters}                          & \textbf{Values}                                \\ \hline
        \cite{lsdnn} Architecture                           & $2 \times k_{on} ;   k_{on} ; 2 \times k_{on}$ \\ \hline
        LC-LSDNN Architecture                        & $k_{on} ;   k_{on}/2 ; k_{on}$                 \\ \hline
        Hidden layer activation function             & ReLU                                           \\ \hline
        Loss Function                                & MSE                                            \\ \hline
        Optimizer                                    & ADAM                                           \\ \hline
        Epochs                                       & 500                                            \\ \hline
        Number of training samples                   & 20000                                          \\ \hline
        Number of testing samples                    & 2000                                           \\ \hline
        \end{tabular}

    }
\end{table}

The DNN is trained using a MATLAB-simulated dataset. Input comprises LS-estimated channel values, while the corresponding channel frequency response (CFR) can be used as labels. Since CFR is not available in real-world systems, we used received signals with very high SNR as training labels~\cite{deepwiphy}. This can be easily obtained by placing the transmitter and receivers close to each other.  
20,000 training samples were generated, with 16,000 samples allocated for the training and 4,000 samples designated for validation. 
Independent testing sets comprising 2,000 samples for each testing SNR were also generated to evaluate the system performance. The training dataset was generated for a fixed SNR of 10dB. The rest of the DNN parameters are given in Table~\ref{tab:dnnParam}.



\subsection{Performance of Floating Point Architectures}
\begin{figure}[!t]
\centering
\includegraphics[scale=0.375]{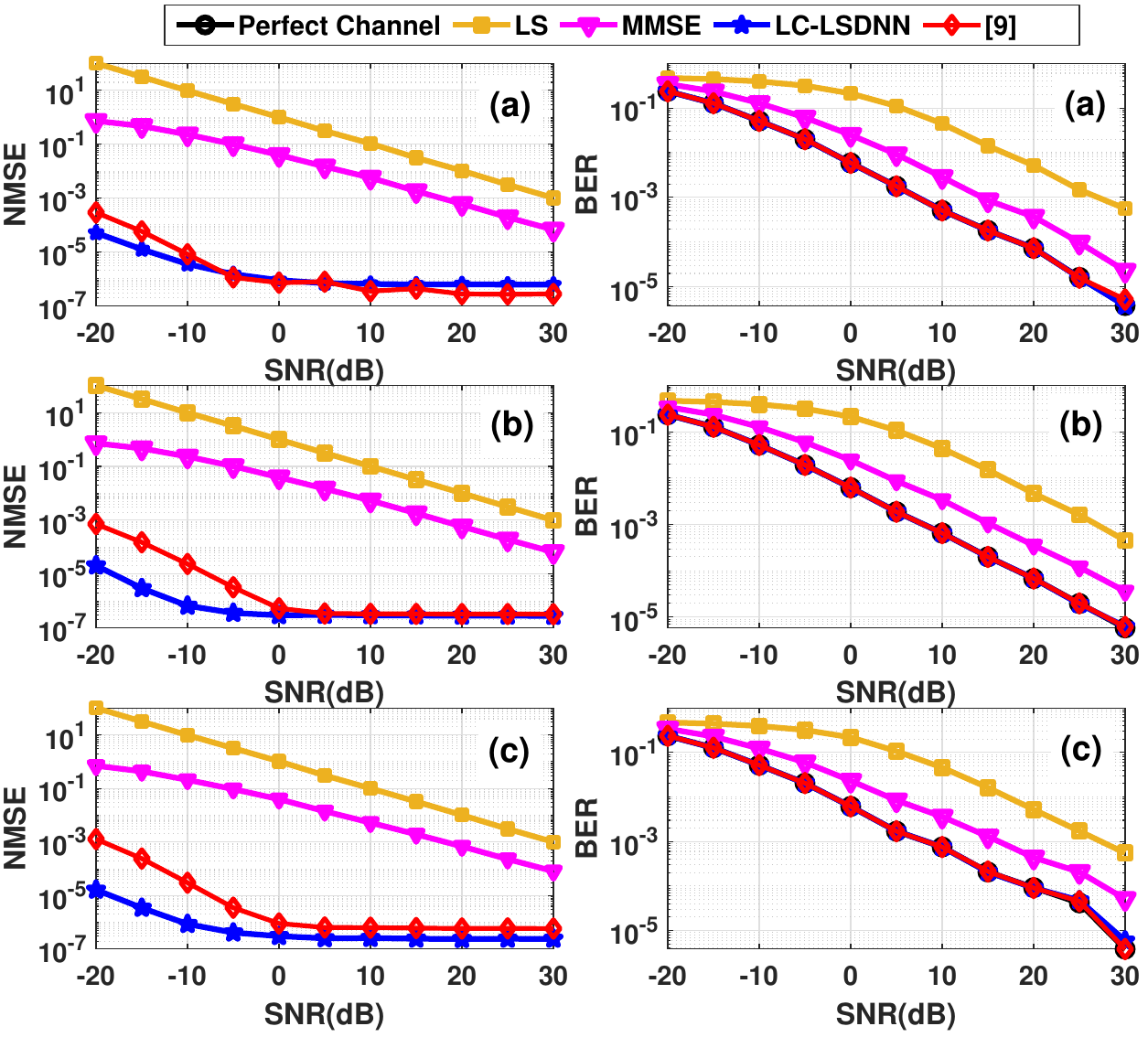}
\caption{\small NMSE  and BER performance comparison of LC-LSDNN with LS, MMSE and \cite{lsdnn} approaches for (a) VTV- urban expressway oncoming, (b)VTV- urban canyon, and (c) RTV- urban canyon channel models.}
    \label{fig:perfComp}
    \vspace{-0.2cm}
\end{figure}

We have validated the functional correctness of the LC-LSDNN over a wide range of SNR under realistic multipath fading environments with zero-doppler conditions. We use well-known channel models \cite{channelmodels} commonly associated with IEEE 802.11p systems, replicating \ac{vtv} and \ac{rtv} environments.
As shown in Fig.~\ref{fig:perfComp}, the NMSE and BER performance improves with SNR for all approaches. The LC-LSDNN offers nearly the same NMSE and BER as that of \cite{lsdnn} for all SNRs and three different channels. Both DL-based approaches outperform the LS and MMSE. Furthermore, DL-based CE offers BER, which is nearly the same as perfect CE, thereby validating their functional superiority. 



\subsection{Theoretical Complexity analysis}
In Table~\ref{tab:complexityComparison}, we compare the LC-LSDNN and \cite{lsdnn} in terms of a number of parameters and \ac{mac} operations. \cite{lsdnn} utilizes a single \ac{dnn} with an input and output layer of size $2 \times k_{on}$ and a hidden layer with a size half of that of the input layer. In contrast, the LC-LSDNN employs two DNNs with identical architecture as in \cite{lsdnn}, featuring an input and output layer size of $k_{on}$ and a hidden layer size again half that of the input layer. 
As shown in Table~\ref{tab:complexityComparison}, LC-LSDNN offers significant savings of 49\%  and 50\% in the number of parameters and MAC operations, respectively, over \cite{lsdnn}. 

\begin{table}[!h]
\centering
\caption{\small Complexity Comparisong of LC-LSDNN and \cite{lsdnn}.}
\label{tab:complexityComparison}
\renewcommand{\arraystretch}{1.2}
 \resizebox{\columnwidth}{!}{
\begin{tabular}{|l|ll|ll|l|l|}
\hline
\multirow{2}{*}{\textbf{}} & \multicolumn{2}{c|}{\textbf{Hidden   Layer}}                                    & \multicolumn{2}{c|}{\textbf{Output   Layer}}                                    & \multicolumn{1}{c|}{\multirow{2}{*}{\textbf{\begin{tabular}[c]{@{}c@{}}Total \\ Parameters\end{tabular}}}} & \multicolumn{1}{c|}{\multirow{2}{*}{\textbf{\begin{tabular}[c]{@{}c@{}}Total  \\ MACs\end{tabular}}}} \\ \cline{2-5}
                           & \multicolumn{1}{c|}{\textbf{\# Weights}} & \multicolumn{1}{c|}{\textbf{\#Bias}} & \multicolumn{1}{c|}{\textbf{\# Weights}} & \multicolumn{1}{c|}{\textbf{\#Bias}} & \multicolumn{1}{c|}{}                                                                                      & \multicolumn{1}{c|}{}                                                                                 \\ \hline
\textbf{\cite{lsdnn}}             & \multicolumn{1}{l|}{$104 \times 52$}     & $52$                                 & \multicolumn{1}{l|}{$52 \times 104$}     & $104$                                & \textbf{10972}                                                                                             & \textbf{10816}                                                                                        \\ \hline
\textbf{LC-LSDNN}        & \multicolumn{1}{l|}{$2(52 \times 26)$}   & $2(26)$                              & \multicolumn{1}{l|}{$2(26 \times 52)$}   & $2(52)$                              & \textbf{5564}                                                                                              & \textbf{5408}                                                                                         \\ \hline
\end{tabular}

}
\vspace{-0.25cm}
\end{table}

%% file: 5_soc_implementation.tex
\section{Hardware Architectures for CE on ZSoC} \label{sec:socImplementation}
The algorithms to architecture mapping of MMSE and LC-LSDNN on the ZSoC are discussed in this section. Other algorithms, such as LS and \cite{lsdnn}, are already realized on ZSoC, and their source codes are publicly available. 











\subsection{LC-LSDNN Architecture}
\begin{figure}[!t]
\centering
\includegraphics[scale=0.55]{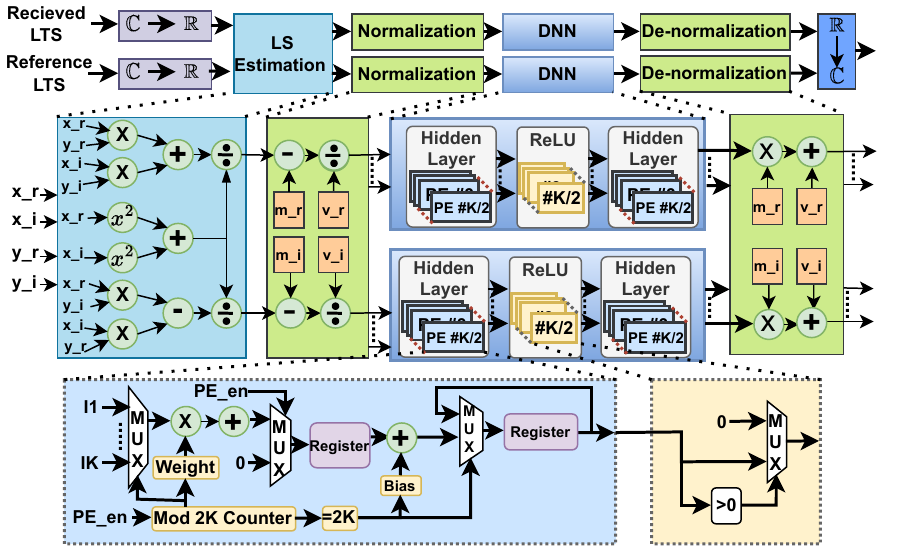}
\caption{\small The architecture of LC-LSDNN.}
    \label{fig:lsdnnArch}
    \vspace{-0.2cm}
\end{figure}

The LC-LSDNN algorithm comprises four operations: LS estimation, normalization, DNN inference, and denormalization. Fig.~\ref{fig:lsdnnArch} illustrates the proposed FPGA architecture for LC-LSDNN realization. The received complex LTS undergo LS estimation in the receiver, which, on hardware, involves six real multiplications, three additions, and two real divisions. Depending on FPGA resources, these operations can be parallelized for subcarriers to optimize the latency. For IEEE 802.11p, BPSK modulated LTS enables simplifying LS to multiplexer operations, as LS output remains unchanged or negated based on the reference LTS being +1 or -1.

The LS output is normalized by subtracting the mean and dividing it by the standard deviation. The mean and standard deviation values are pre-calculated during the DNN training and are pre-stored in the internal memory. The DNN processes the normalized LS  estimates. In LC-LSDNN, we use different DNNs for real and imaginary components of the input samples. The fundamental computational unit within the DNN is a neuron or \ac{pe}. Each \ac{pe} executes a \ac{mac} operation on inputs, employing their associated weights. These inputs are sourced from the previous layer of the network.
After the  \ac{mac}  operation, the bias is added, and the resulting output is processed by the \ac{relu} activation function. The  \ac{relu} decides between assigning zero or retaining the PE's output, depending on whether the output is negative or positive respectively. The schematic depiction of a  \ac{pe}'s architectural design is provided in Fig.~\ref{fig:lsdnnArch}. 

All the \ac{pe}s within a layer are implemented in parallel. At the same time, the operations within each \ac{pe} are pipelined to achieve a balanced trade-off between resource utilization and latency.
The output of the \ac{dnn} is denormalized to obtain the final output. The outputs from both \ac{dnn}s are combined to obtain complex samples, which are then used for channel equalization.

\subsection{MMSE Architecture} \label{sec:socArch_mmse}
\begin{figure}[!t]
\centering
\includegraphics[scale=0.65]{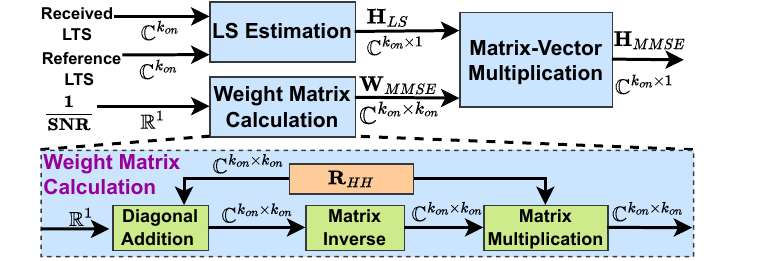}
\caption{\small \ac{mmse} Architecture.}
    \label{fig:mmseArch}
    \vspace{-0.2cm}
\end{figure}

As discussed in \cite{lsdnn}, the MMSE is computationally complex and from Eq.~\ref{eqn:mmse}, we can observe that the arithmetic operations in MMSE are not hardware-friendly. The architecture of the MMSE involves the LS estimation followed by multiplication with weight matrix as shown in Fig.~\ref{fig:mmseArch}. The calculation of the MMSE weight matrix involves the addition of the inverse of the SNR to the diagonal elements of the channel auto-correlation matrix \(R_{HH}\), followed by matrix inverse operation of size  \(k_{on} \times k_{on}\). The resulting inverse matrix is then multiplied with the channel auto-correlation matrix to obtain the MMSE weight matrix.


Compared to existing works, we focus on comparing the complexity of MMSE using different matrix inversion approaches. The matrix inversion through the method of cofactors is unfeasible due to its excessively high computation time. In this work, we have designed three \ac{mmse} architectures using three matrix inversion approaches on ZSoC: 1) Gauss-Jordan \cite{MMSE_GJ}, 2) QR\cite{MMSE_QR}, and 3) LU decomposition \cite{MMSE_LU}. The hardware architectures of these MMSE algorithms are designed using reference matrix inversion examples provided by AMD-Xilinx. The functional and complexity performance of these architectures is discussed in Section~\ref{sec:results}.

\begin{table*}[!b]
\centering
\caption{\small Complexity comparison of various CE approaches for SPFL WL and SDMA with fixed 50 MHz clock frequency.}
\label{tab:compLsdnn}
 \resizebox{\textwidth}{!}{

\begin{tabular}{|cc|c|c|c|c|c|c|c|c|}
\hline
\multicolumn{2}{|c|}{\textbf{Architectures}}                       & \textbf{Interface} & \textbf{Execution time(us)} & \textbf{LUT}   & \textbf{FF}    & \textbf{BRAM} & \textbf{DSP} & \textbf{Total Power(W)} & \textbf{PL Power (W)} \\ \hline
\multicolumn{2}{|c|}{\textbf{LS}}                                  & \textbf{SDMA}            & 13.3                        & 14721          & 15094          & 14            & 96           & 1.969                   & 0.438                 \\ \hline
\multicolumn{1}{|c|}{\multirow{3}{*}{\textbf{MMSE}}} & \textbf{GJ} & \textbf{SDMA}            & 25859              & 32621          & 24917          & 124           & 362          & 2.479                   & 0.948                 \\ \cline{2-10} 
\multicolumn{1}{|c|}{}                               & \textbf{QR} & \textbf{SDMA}            & 23041                       & 37497          & 29019          & 149           & 383          & 2.193                   & 0.663                 \\ \cline{2-10} 
\multicolumn{1}{|c|}{}                               & \textbf{LU} & \textbf{SDMA}            & 21359                       & 43356          & 32870          & 181  & 522 & 2.711          & 1.18         \\ \hline
\multicolumn{2}{|c|}{\multirow{2}{*}{\cite{lsdnn}}}              & \textbf{SDMA}            & 27.99                       & 41224          & 37956          & 79.5            & 208          & 2.593                   & 1.063                 \\ \cline{3-10} 
\multicolumn{2}{|c|}{}                                             & \textbf{MM}            & 26.84                       & 39285          & 35622          & 79.5            & 208          & 2.204                   & 0.674                 \\ \hline
\multicolumn{2}{|c|}{\multirow{2}{*}{\textbf{LC-LSDNN}}}           & \textbf{SDMA}            & 17.46                      & 38589 & 37779 & 35            & 184          & 2.128                   & 0.598                 \\ \cline{3-10} 
\multicolumn{2}{|c|}{}                                             & \textbf{MM}            & 17.07                        & 38670          & 37618          & 38            & 184          & 2.152                   & 0.622                 \\ \hline
\end{tabular}

}

\end{table*}

\begin{figure}[!b]
\vspace{-0.25cm}
\centering
\includegraphics[scale=0.57]{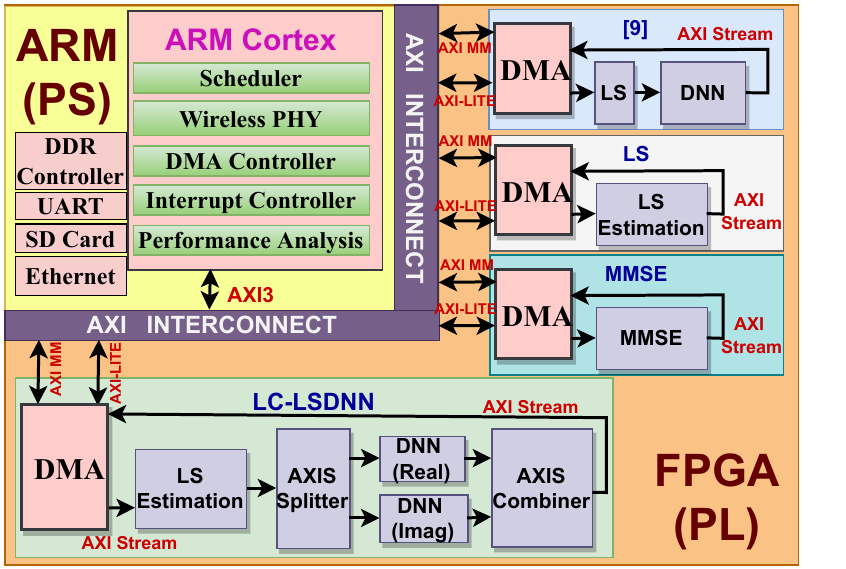}
\caption{\small Hardware-software co-design architecture of CE on ZSoC.}
    \label{fig:sysDesign}
\end{figure}

\subsection{System Level Architecture for CE on ZSoC}
We have realized various CE approaches on ZSoC comprising dual core Cortex-A9 processor from ARM as processing system (PS) and ultra-scale FPGA from AMD-Xilinx as programmable logic (PL). The corresponding architecture obtained via hardware-software co-design is shown in Fig.~\ref{fig:sysDesign}. All CE hardware IPs are designed with an advanced extensible interface (AXI) stream interface. These IPs can be configured by the processor using AXI-Lite interface, and data communication with memory is done using AXI direct memory access (AXI DMA). In LC-LSDNN, AXI broadcaster IP is used to split the complex estimates into their real and imaginary components before feeding them to their respective DNNs, and the AXI combiner is employed to merge their outputs before transmitting them to the PS for further processing. The remainder of the wireless PHY is implemented in the PS, along with the DMA controller, interrupt controller, and scheduler. 


%% file: 6_results.tex
\section{Performance and Complexity Analysis} \label{sec:results}

\begin{figure}[!h]
\centering
\includegraphics[scale=0.33]{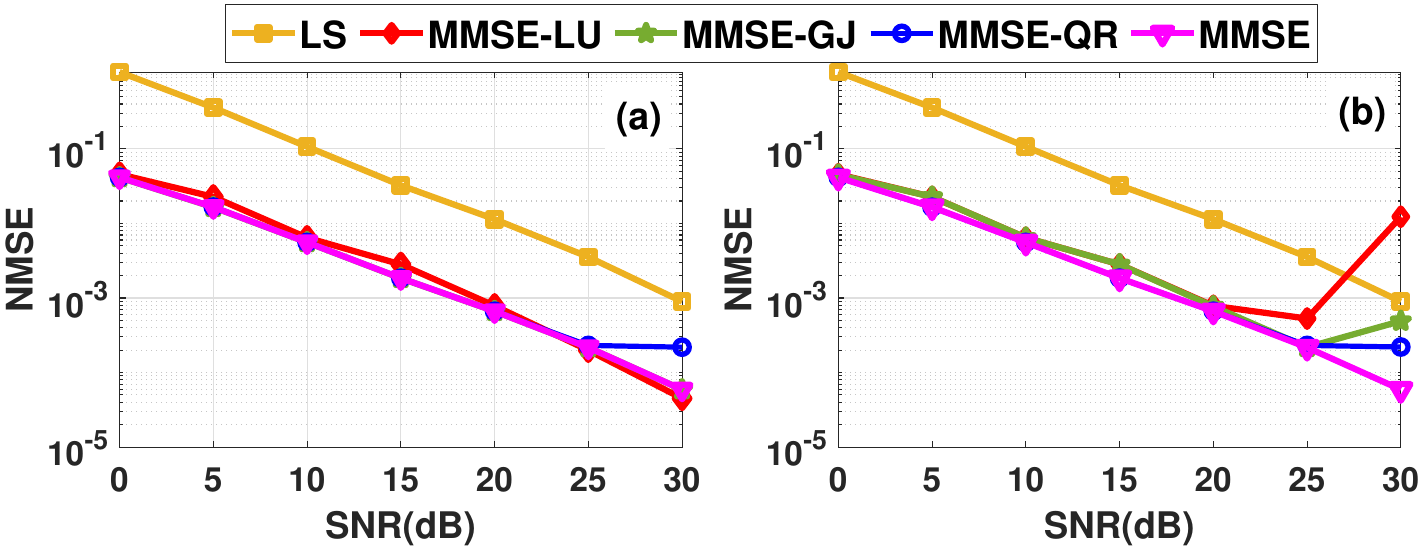}
\caption{\small Double precision and Single precision floating point implementation of MMSE channel estimation schemes.}
    \label{fig:mmseWl}
\end{figure}

In this section, we present functional performance and complexity results on the ZC706 platform, comparing them with state-of-the-art CE approaches. We evaluate three matrix inversion algorithms for MMSE, as detailed in Section~\ref{sec:socArch_mmse}. Figure~\ref{fig:mmseWl}(a) illustrates that MMSE outperforms LS across all SNRs for DPFL WL. However, the performance of MMSE-QR degrades slightly at higher SNR due to erroneous matrix inverse operations. In the case of SPFL WL in Fig.~\ref{fig:mmseWl}(b), the performance of all MMSE hardware IPs degrades at high SNR. Hence, we do not explore fixed-point architectures for MMSE and use DPFL WL for the complexity analysis.

Since the data to be processed by the receiver PHY is buffered in memory, the interface and type of communication also impact the latency performance. We have considered three different types of interfaces: 1) Simple direct memory access (SDMA), 2) Scatter Gather DMA (SGDMA), and 3) Memory-mapped (MM). All these approaches are based on the AXI protocol. In SDMA and SGDMA, the DMA is used to read and write the data between memory and hardware IP, while MM allows the hardware IP to directly read and write the data from memory. Among DMAs, SDMA can access contiguous data, while SGDMA can access non-contiguous data. Thus, SDMA needs frequent configuration for non-contiguous data, making SGDMA a preferred choice.


\begin{table*}[!h]
\centering
\caption{\small Complexity comparison of various CE approaches using SDMA and SGDMA with architecture operating at the maximum clock frequency.}
\label{tab:clockFreq}
\renewcommand{\arraystretch}{1.05}
 \resizebox{\textwidth}{!}{

\begin{tabular}{|cc|c|c|c|c|c|c|c|c|c|}
\hline
\multicolumn{2}{|c|}{\textbf{Architectures}}                                        & \textbf{Max. clock   frequency}    & \textbf{Interface} & \textbf{Execution   time(us)} & \textbf{LUT}   & \textbf{FF}    & \textbf{BRAM} & \textbf{DSP} & \textbf{Total Power(W)} & \textbf{PL Power (W)} \\ \hline
\multicolumn{1}{|c|}{\multirow{6}{*}{\textbf{MMSE}}} & \multirow{2}{*}{\textbf{LU}} & \multirow{2}{*}{\textbf{83.3}}   & \textbf{SDMA}            & 13366                         & 43797          & 40332          & 183           & 515          & 3.082                   & 1.557                 \\ \cline{4-11} 
\multicolumn{1}{|c|}{}                               &                              &                                  & \textbf{SGDMA}            & 13376                         & 44148          & 41845          & 183           & 515          & 3.196                   & 1.671                 \\ \cline{2-11} 
\multicolumn{1}{|c|}{}                               & \multirow{2}{*}{\textbf{GJ}} & \multirow{2}{*}{\textbf{111.11}} & \textbf{SDMA}            & 13811                         & \textbf{32505} & \textbf{34471} & 126           & 358          & 3.356                   & 1.833                 \\ \cline{4-11} 
\multicolumn{1}{|c|}{}                               &                              &                                  & \textbf{SGDMA}            & 13796                         & 34203          & 37198          & 126           & 358          & 3.344                   & 1.821                 \\ \cline{2-11} 
\multicolumn{1}{|c|}{}                               & \multirow{2}{*}{\textbf{QR}} & \multirow{2}{*}{\textbf{111.11}} & \textbf{SDMA}            & 12045                         & 37795          & 38426          & 151           & 395          & 2.746                   & 1.225                 \\ \cline{4-11} 
\multicolumn{1}{|c|}{}                               &                              &                                  & \textbf{SGDMA}            & 12006                         & 39435          & 41153          & 151           & 395          & 2.758                   & 1.237                 \\ \hline
\multicolumn{2}{|c|}{\multirow{2}{*}{\cite{lsdnn}}}                               & \multirow{2}{*}{\textbf{125}}    & \textbf{SDMA}            & 14.67                         & 35193          & 38945          & 83            & 138          & 2.164          & 0.634        \\ \cline{4-11} 
\multicolumn{2}{|c|}{}                                                              &                                  & \textbf{SGDMA}            & 12.5                          & 36786          & 41672          & 83            & 138          & 2.593                   & 1.072                 \\ \hline
\multicolumn{2}{|c|}{\multirow{3}{*}{\textbf{LC-LSDNN}}}                            & \textbf{125}                     & \textbf{SDMA}            & 11.144                         & 36827          & 47371          & \textbf{35}   & 116 & 2.569                   & 1.048                 \\ \cline{3-11} 
\multicolumn{2}{|c|}{}                                                              & \multirow{2}{*}{\textbf{200}}    & \textbf{SDMA}            & 7.41                          & 58387          & 89137          & \textbf{35}   & \textbf{112} & 4.217                   & 2.681                 \\ \cline{4-11} 
\multicolumn{2}{|c|}{}                                                              &                                  & \textbf{SGDMA}            & \textbf{5.36}                 & 60016          & 91896          & \textbf{35}   & \textbf{112} & 4.286                   & 2.75                  \\ \hline
\end{tabular}

}

\end{table*}

\begin{table*}[!h]
\centering
\caption{\small Complexity comparison of fixed-point LC-LSDNN and \cite{lsdnn} with SGDMA at maximum clock frequency.}
\label{tab:clockFreq}
\renewcommand{\arraystretch}{1.1}
 \resizebox{\textwidth}{!}{
\begin{tabular}{|l|l|l|l|l|l|l|l|l|l|}
\hline
\textbf{Architectures} & \textbf{Word Length}                                                   & \textbf{Max. clock   frequency} & \textbf{Execution   time(us)} & \textbf{LUT}   & \textbf{FF}    & \textbf{BRAM} & \textbf{DSP} & \textbf{Total Power(W)} & \textbf{PL Power (W)} \\ \hline
\cite{lsdnn}         & \textbf{\textless{}24,8\textgreater ,   \textless{}18,2\textgreater{}} & 125                           & 6.844                         & \textbf{28801} & \textbf{30170} & \textbf{31}   & \textbf{169} & \textbf{2.548}          & \textbf{1.027}        \\ \hline
\textbf{LC-LSDNN}      & \textbf{\textless{}24,8\textgreater ,   \textless{}18,2\textgreater{}} & 200                           & 2.52                          & 86165          & 89244          & 9             & 172          & 5.764                   & 4.228                 \\ \hline
\end{tabular}

}

\end{table*}

In Table~\ref{tab:compLsdnn}, we compare four CE approaches in terms of FPGA resource utilization, execution time, and power consumption. We assume the input data is in contiguous memory and, hence, limit our discussion to SDMA and MM interfaces. Also, all architectures are designed to work at a fixed clock frequency of 50 MHz. Among conventional statistical approaches, LS is efficient but performs poorly compared to MMSE in CE functionality. In the case of MMSE, MMSE-LU demonstrates the fastest execution time with a slight increase in resources and power consumption. On the contrary, the MMSE-GJ offers the highest execution time with lower resources and power consumption, among other MMSE approaches. Though MMSE offers better performance than LS, there is a need for alternative approaches that can offer performance that is the same or better than MMSE with comparable complexity as that of LS. 

Next, we compare the LS with \cite{lsdnn} and proposed LC-LSDNN.
Please note that the results for LC-LSDNN are for both the real and imaginary DNNs.
For SDMA, LC-LSDNN offers a 37.6\% increase in speed, accompanied by a substantial 55.97\% reduction in BRAM utilization, an 11.5\% reduction in DSP utilization, and 17.93\% lower power consumption than \cite{lsdnn}. 
Both DL-based approaches significantly outperform MMSE in functional and complexity comparison. This is mainly due to replacing complex MMSE signal processing operations with hardware-friendly DL arithmetic operations. 
By removing DMA and using the MM interface, the execution time of LC-LSDNN is reduced by 2.2\% compared to the SDMA and by 36.4\% compared to the \cite{lsdnn} with further reduction in resource utilization due to the absence of DMA, as LC-LSDNN can use two parallel interfaces for memory read and write compared to a single DNN in \cite{lsdnn}. 
The LC-LSDNN has only 25\% higher execution time than LS compared to 100\% higher execution time in LSDNN.

Next, we optimized the critical path and increased the clock frequency at which the architecture operates without losing functional accuracy. As shown in Table~\ref{tab:clockFreq}, the maximum clock frequency is different for each architecture, and LC-LSDNN can work at highest frequency due to a smaller critical path.

Among the MMSE variants, MMSE-LU had the longest critical path, resulting in the smallest attainable clock frequency of 83.3 MHz. In contrast, MMSE-GJ and MMSE-QR achieved higher frequencies of 111.11 MHz. These adjustments significantly reduced execution times, with decreases of 46\%, 47.7\%, and 37.4\% for MMSE-GJ, MMSE-QR, and MMSE-LU, respectively. Notably, with the higher clock periods, MMSE-QR achieved the shortest execution time. 

 Compared to 125 MHz in \cite{lsdnn}, LC-LSDNN can operate at a much higher clock frequency of 200 MHz. At this frequency, LC-LSDNN's execution time is reduced to $5.36 \mu s$, which is 57\% faster than \cite{lsdnn} at 125 MHz and 59.7\% faster than LS at 50 MHz. However, this improvement in latency is accompanied by an increase in power dissipation, with LC-LSDNN consuming 39.5\% more power than \cite{lsdnn}. Even when operated at the same clock frequency of 125 MHz, LC-LSDNN is 13\% faster than \cite{lsdnn}, highlighting the low complexity and high-speed characteristics of the LC-LSDNN. 

We optimized the architectures further using fixed-point WL. Our detailed study identified a minimum WL of 24 bits, with 8 bits for the integer part and 16 bits for the fractional part, denoted as $<24,8>$, ensuring computational accuracy. Additionally, a WL of $<18,2>$ proved sufficient for DNN parameters. With the reduced WL, the execution time of LC-LSDNN was further reduced by 51\%, and BRAM utilization decreased by 74.2\%. However, DSP utilization increased by 34.8\%, due to further possibility of parallel operations. Depending on the desired execution time, we can reduce resource utilization by serializing the architecture.

%% file: 7_conclusion.tex
\section{Conclusion} \label{sec:conclusions}
We developed a novel low complexity high speed deep neural network augmented least square (LC-LSDNN) based channel estimation (CE) and efficiently mapped it on zynq system on chip (ZSoC). The novelty of the LC-LSDNN is to use different DNNs for real and imaginary values of received complex symbols. This helps reduce the size
of DL by around 50\% and optimize the critical path, allowing it to operate at 60\%  higher clock frequency. In addition, we offer significant savings in resource utilization and power consumption over state-of-the-art DL-based CE without compromising on functional accuracy. 

